\documentclass[conference]{IEEEtran}
\makeatletter
\def\ps@headings{%
\def\@oddhead{\mbox{}\scriptsize\rightmark \hfil \thepage}%
\def\@evenhead{\scriptsize\thepage \hfil \leftmark\mbox{}}%
\def\@oddfoot{}%
\def\@evenfoot{}}
\makeatother 

\usepackage{graphicx}
\usepackage{tabularx}
\usepackage{times}
\usepackage{amsmath}
\usepackage{amssymb}
\usepackage{cite}
\usepackage{epsfig}
\usepackage{subfigure}
\usepackage[ruled]{algorithm2e}

\newtheorem{definition}{{\bf Definition}}

\begin{document}
\title{Real-Time Misbehavior Detection in IEEE 802.11e Based WLANs}

\author{\IEEEauthorblockN{{Xianghui Cao\IEEEauthorrefmark{1},
Lu Liu\IEEEauthorrefmark{1}, Wenlong Shen\IEEEauthorrefmark{1}, Jin Tang\IEEEauthorrefmark{2}  and
Yu Cheng\IEEEauthorrefmark{1}}
\IEEEauthorblockA{\IEEEauthorrefmark{1}Department of Electrical and
Computer Engineering, Illinois Institute of Technology, USA
\\Email: xcao10@iit.edu; \{lliu41,wshen7\}@hawk.iit.edu; cheng@iit.edu
}
\IEEEauthorblockA{\IEEEauthorrefmark{2}AT\&T Labs, USA. Email: jin.tang@att.com}
}}

\maketitle

\begin{abstract}
The Enhanced Distributed Channel Access (EDCA) specification in the IEEE 802.11e standard supports heterogeneous backoff parameters and arbitration inter-frame space (AIFS), which makes a selfish node easy to manipulate these parameters and misbehave. In this case, the network-wide fairness cannot be achieved any longer. Many existing misbehavior detectors, primarily designed for legacy IEEE 802.11 networks, become inapplicable in such a heterogeneous network configuration. In this paper, we propose a novel real-time hybrid-share (HS) misbehavior detector for IEEE 802.11e based wireless local area networks (WLANs). The detector keeps updating its state based on every successful transmission and makes detection decisions by comparing its state with a threshold. We develop mathematical analysis of the detector performance in terms of both false positive rate and average detection rate. Numerical results show that the proposed detector can effectively detect both contention window based and AIFS based misbehavior with only a short detection window.
\end{abstract}


\IEEEpeerreviewmaketitle

\section{Introduction}\label{sec:introduction}
To support rapid growing applications (especially multimedia ones) of wireless local area networks (WLANs), the IEEE 802.11e standard adopts the Enhanced Distributed Channel Access (EDCA) mechanism to provide media access control (MAC) level differentiation in quality of service (QoS) \cite{zhao2013scalable,chendeb2014towards}. With EDCA, network traffic is prioritized and classified into several access categories (ACs). Service differentiation is realized by assigning different parameters for each AC, including the minimum and maximum contention window sizes (CWmin and CWmax, respectively), the arbitration inter-frame space (AIFS) number and transmission opportunity (TXOP) limit \cite{standards2005wireless}.

In IEEE 802.11e based WLANs, a selfish/misbehaving node can deliberately manipulate those parameters to gain advantage over others. For example, it can use a smaller AIFS to wait for shorter time than others in the same AC before accessing the medium. As a result, it can access the medium more frequently, and hence gain a higher priority for data transmission. It is even possible for an intensively misbehaving node to block the transmissions from other nodes and cause the so-called denial of service attack. Therefore, real-time misbehavior detection is demanded in order to isolate such a node and alleviate its impact to the network.

Due to the random access which is based on the carrier sense and multiple access with collision avoidance (CSMA/CA) MAC protocol, we usually need to monitor each node for a period of time to judge whether it is misbehaving or not. Since it is difficult to extract necessary information from collided transmissions, information conveyed in successful transmissions is perhaps the only measurement can be utilized for detection. Toledo \emph{et al.} proposed to detect backoff misbehavior by checking whether the idle time between consecutive successful transmissions from a target node obeys the normal distribution \cite{lopez2007robust}. Exploiting the fairness property across the network, Tang \emph{et al.} designed a light-weight fair-share detector, which does not rely on the idle time distribution \cite{tang2014real}. However, with multiple ACs in an IEEE 802.11e WLAN, the network-wide fairness as achieved in legacy IEEE 802.11 based networks dose not hold any longer \cite{bianchi2005understanding}, making the above detectors generally inapplicable.

To detect backoff misbehavior in IEEE 802.11e networks, Szott \emph{et al.} proposed a $\chi^2$ detector by comparing the measured and expected backoff values \cite{szott2011detecting}. However, the exact values of backoff periods followed by unsuccessful transmissions may be hard to measure. The detector in \cite{serrano2010detecting}, however, takes advantage of the fact that the interval between two consecutive successful transmissions is uniformly distributed in [0,~CWmin) providing that the packet in the second transmission was not retransmitted before. Nevertheless, the detector delay could be very high. While there are works well addressed the TXOP misbehavior \cite{ahn2011fair}, efficient and real-time detection of contention window and AIFS misbehavior still remains open.

In this paper, we propose a new detector to deal with misbehavior in IEEE 802.11e networks. We focus on both contention window and AIFS misbehavior. The major contributions in this paper can be summarized as follows. We analyze the misbehavior strategy in IEEE 802.11e networks and show that a selfish node can gain significant advantage over other nodes by manipulating its contention window or AIFS. We also demonstrate that the existing fair-share based detector for legacy IEEE 802.11 networks is unable to detect certain misbehavior in the IEEE 802.11e cases with multiple priority classes. Then, we propose a mathematical model of the percentage of resource sharing for a node in each priority class. Based on this, we design a novel hybrid-share detector and develop analytical results of the detector performance in terms of false positive rate and average detection rate. We also present numerical results to demonstrate the performance in various aspects including different threshold, misbehaving intensity and detection window. The remainder of this paper is organized as follows. Section \ref{sec:overview} overviews the problem. Following the mathematical MAC model in Section \ref{sec:model}, our detector is designed and evaluated in Section \ref{sec:hsdetector}. Numerical results are presented in Section \ref{sec:numerical} and the paper is concluded in Section \ref{sec:conclusion}.

\section{Problem Overview} \label{sec:overview}

\subsection{IEEE 802.11e EDCA}
In IEEE 802.11 based wireless local area networks (WLANs), the channel access among nodes is coordinated by the CSMA/CA mechanism. Time is divided into equal slots. Before transmission, a node should sense the medium idle until a backoff timer expires. Each node takes the binary exponential backoff strategy to access the channel with the backoff timer at each backoff stage initialized at a value randomly chosen from $[0,CW-1]$. The contention window size $CW$ is initialized at CWmin and doubles (until CWmax) once a transmission is unsuccessful (a packet will be retransmitted at most for a certain number of times). Once the medium is busy, the backoff timer will be suspended until it becomes idle again. The CSMA/CA mechanism also uses an inter-frame space time to defer a transmission or backoff period in order to give way to high priority messages. Unlike the distributed coordination function (DCF) mechanism in legacy IEEE 802.11 standard, the Enhanced Distributed Channel Access (EDCA) specification in IEEE 802.11e supports hybrid backoff parameters and arbitration inter-frame space (AIFS). In default, there are four priority classes (access categories) defined in IEEE 802.11e EDCA \cite{standards2005wireless}, as shown in Table \ref{Table:EDCA}.
\begin{table}[ht]
\centering
\label{Table:EDCA}
\caption{EDCA default settings.}
\begin{tabular}{|c|c|c|c|}
  \hline
    \textbf{Access category}	& \textbf{CWmin}	& \textbf{CWmax}	& \textbf{AIFSN} \\
  \hline
    Background $\rm{AC}_{\rm{BK}}$	& {aCWmin}	& {aCWmax}	& 7 \\
    Best Effort $\rm{AC}_{\rm{BE}}$ & {aCWmin} & {aCWmax}	& 3 \\
    Video $\rm{AC}_{\rm{VI}}$	& ({aCWmin}+1)/2-1	& {aCWmin}	& 2 \\
    Voice $\rm{AC}_{\rm{VO}}$	& ({aCWmin}+1)/4-1	& ({aCWmin}+1)/2-1	& 2 \\
  \hline
\end{tabular}
\vspace{-2mm}
\end{table}

In this paper, we consider the general cases that there are $c$ priority classes, each of which is assigned contention window sizes CWmin$_i$ and CWmax$_i$, and inter-frame space AIFS$_i$=AIFSN$_i$*aSlotTime+aSIFSTime, where AIFSN is the number of slots after a short inter-frame space duration a node should defer before either invoking a backoff or starting a transmission. The parameters are assigned by the AP.

\subsection{Misbehavior Analysis}
\begin{figure}[htbp]
\vspace{-3mm}
  \centering
  \includegraphics[width=2.3in]{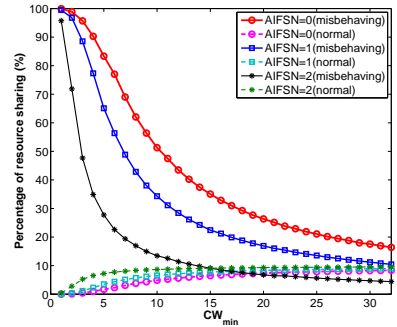}
\vspace{-4mm}
  \caption{Impact of MAC misbehavior.}\label{fig:CWvsAIFSN}
\vspace{-3mm}
\end{figure}

A misbehaving node may use different parameters other than those assigned by the AP, to gain a higher sharing of the resource. Fig. \ref{fig:CWvsAIFSN} illustrates the impact of a misbehaving node and shows the percentages of resource sharing of the misbehaving node and a normal node. Here, the percentage of resource sharing is defined as the portion of throughput contribution from a particular node over the total network throughput. In this figure, we consider a network consisting of 10 normal nodes and one misbehaving node. Each node always has packets in its buffer for sending out. Each normal node takes MAC parameters as $\textrm{CWmin}=15,\textrm{CWmax}=1023$ and $\textrm{AIFSN}=2$, while the misbehaving node takes $\textrm{CWmin}=1\sim 32,\textrm{CWmax}=1023$ and $\textrm{AIFSN}=0\sim 2$.

The figure clearly demonstrates that the misbehaving node can gain significant advantage over the other nodes by manipulating its MAC parameters. Moreover, the impacts of CWmin and AIFSN are different. For example, in order to achieve 10\% more throughput, the misbehaving node needs to reduce its CWmin to a much smaller value (e.g., from 15 to less than 7); while, this can also be achieved by simply reducing its AIFSN from 2 to 1. In other words, the misbehaving impact on the network is more sensitive to AIFSN than CWmin.

In this paper, we consider both CWmin and AIFSN misbehavior. While the proposed detector is also effective for CWmax misbehavior, a malicious node may prefer to manipulate CWmin over CWmax since the former strategy has greater impact. We focus on saturated traffic case, i.e., each node always has packets for transmitting to the AP. Otherwise, for a light-loaded network, a misbehaving node may not have much impact on the other normal nodes. Our problem is to detect such misbehavior at the AP in real-time.

\subsection{Fair-Share Detector and Challenges}
Consider a WLAN consisting of one AP and $n$ nodes which locate inside each other's communication range. The nodes compete for accessing a channel and sending packets to the AP. In the legacy IEEE 802.11 standard, the DCF mechanism guarantees that each node will share the same portion of the channel resource and maintains fairness across the network. For an arbitrary node $v$, let a binary variable $I_v$ be the indicator of whether a packet received by the AP is from node $v$ or not. In normal cases, due to the network-wide fairness guarantee, we have probability $\mathbb{P}[I_v=1] = \frac1n$.

A misbehaving node can gain unfair share of the resource by manipulating its backoff parameters, e.g, using a smaller CWmin. If the AP records all the received packets, it can notice that more packets are from the selfish one, i.e., $\mathbb{P}[I_v=1] > \frac1n$ if $v$ is misbehaving. In \cite{tang2014real}, we take advantage of this important feature to design a nonparametric cumulative sum (CUSUM) based fair-share misbehavior detector (called FS detector) to detect such misbehavior in real-time, which is described as follows.

For a target node $v$, let $X_k$ be the state of the detector for $v$. $X_k$ initializes at $0$, i.e., $X_0=0$. For the $k$th packet received by the AP, the state of the detector is updated as follows.
\begin{equation}\label{eq:fsdetector}
    X_{k+1} = \left[X_k+(nI_k-1)\right]^+,
\end{equation}
where $x^+=x$ if $x>0$ and $0$ otherwise. In other words, if the packet is from $v$, we have $I_k=1$ and $X_{k+1} = X_k+n-1$; otherwise, $I_k=0$ and $X_{k+1} = X_k-1$. The idea behind is that, due to fair sharing, the nodes roughly take turns to transmit packets. Therefore, the detector state $X_k$ is likely to be bounded. In presence of misbehaving nodes, since $\mathbb{P}[I_v=1] > \frac1n$, the unfair portion of channel sharing will accumulate such that the state of the FS detector associated with each misbehaving node finally becomes unbounded. Thus, we can employ a detection threshold $h$ to decide whether $v$ is misbehaving (i.e., $\delta_k=1$) or not (i.e., $\delta_k=0$) as follows.
\begin{equation}\label{eq:deltakfs}
    \delta_k = \left\{\begin{array}{ll}
        1, \quad & \textrm{if}~X_k \geq h,\\
        0, \quad & \textrm{otherwise}.
    \end{array}\right.
\end{equation}

Because every received packet is counted by the AP in making the detection decisions, with satisfactory accuracy, the FS detector can identify the misbehaving node much faster than many existing detectors. Moreover, the FS detector is nonparametric and lightweight in terms of computation complexity, it is thereby able to provide real-time misbehavior detection services \cite{tang2014real}.

Because the underlying assumption of network-wide fairness no longer holds in the EDCA situation, the FS detector cannot be applied directly in networks with hybrid priority classes. However, since sub-network fairness can be still achieved among the nodes in the same class, a direct extension of the FS detector is to design a distinct one for each class. Specifically, for a node in class $i$, the associated detector should use the number of nodes in this class other than a common $n$ across the whole network as in (\ref{eq:fsdetector}). Nevertheless, such extended FS detector encounters two challenges:
\begin{itemize}
\item If there is only one node in a class, its misbehavior may not be detected. To see this, substituting $n=1$ into (\ref{eq:fsdetector}) we can obtain that $X_{k+1}=[X_k-(1-I_k)]^+\equiv 0$ if $X_0=0$. As a result, $\delta_k\equiv 0$.
\item If all the nodes in one class misbehave, some or all of them may not be detected. Specifically, if they use the same manipulated MAC parameters, the detector sees that none of them is misbehaving; otherwise, at least the one with the lowest throughput among them will be considered as a normal node.
\end{itemize}

Therefore, to overcome the shortcomings of the FS detector, only considering a single priority class is not enough. In the following, we propose a novel hybrid-share detector based on the following analytical MAC model.

\section{MAC Analytical Model}\label{sec:model}
For an arbitrary node $v$ in class $i$, denote $s_i$ as its percentage of resource sharing. In this section, we propose a model for calculating $s_i$. We assume there are $c$ priority classes; each contains $n_i$ nodes which compete for channel access using parameters $W_{i}=\textrm{CWmin}_i$, $\textrm{CWmax}_{i}=2^{m_i}(W_{i}+1)-1$ and AIFS$_i$, where $1\leq i\leq c$ and $m_i=\log_{2}\frac{\textrm{CWmax}_{i}+1}{\textrm{CWmin}_{i}+1}$ is the maximum backoff stage. Here, for simplicity, we assume the maximum retransmission limit for node $v$ is the same as $m_i$\footnote{In general cases, the hybrid share model will be slightly more complicated, but our modeling methodology and the designed detector are still valid.}. Thus, the contention window size of this node in its $j$th backoff stage is
\begin{equation}
    W_{i,j} = 2^j(W_{i}+1)-1.
\end{equation}

Let $p_i$ be the frame blocking probability, i.e., the probability that node $v$ senses a busy channel (and thereby suspends its backoff timer countdown) in a generic slot. According to \cite{xiao2005performance}, the transmitting probability of node $v$ in a generic slot can be calculated as
\begin{align}\label{eq:taui}
\nonumber \tau_i =&\; \frac{1-p_i^{m_i+1}}{(1-p_i)\sum^{m_i}_{j=0}p_i^j
                    \left[1+\frac{1}{1-p_{i}}\sum^{W_{i,j}}_{k=1}\frac{W_{i,j}-k}{W_{i,j}}\right]}\\
\nonumber  =&\; \frac{1-p_i^{m_i+1}}{\sum^{m_i}_{j=0}p_i^j\left(1-p_i+\frac{W_{i}}{2}\right)}\\
           =&\; \frac{2(1-p_i)(1-2p_i)}{(1-2p_i)^2+(W_{i}+1)(1-p_i)\frac{1-(2p_i)^{m_i+1}}{1-p_i^{m_i+1}}}.
\end{align}

Let $\Delta A_i = \textrm{AIFSN}_i-\textrm{AIFSN}_{\min}$ where $\textrm{AIFSN}_{\min}=\min\{\textrm{AIFSN}_j|j=1,\ldots,c\}$. Due to differentiation in AIFS, a node of low priority must wait a longer idle time than a high-priority node after a busy channel period before resuming its backoff timer countdown. Therefore, according to \cite{kosek2011simple}, the frame blocking probability of node $v$ can be calculated as follows.
\begin{align}\label{eq:pi}
\nonumber  p_i = &\; 1-\left[(1-\tau_i)^{n_i-1}\prod^{c}_{k=1,k\neq i}(1-\tau_k)^{n_k}\right]^{\Delta A_i+1}\\
\nonumber  = &\; 1-\left[\frac{1}{1-\tau_i}\prod^{c}_{k=1}(1-\tau_k)^{n_k}\right]^{\Delta A_i+1}\\
        = &\; 1-\left(\frac{1-p_b}{1-\tau_i}\right)^{\Delta A_i+1},
\end{align}
where
\begin{equation}
    p_b=1-\prod^{c}_{k=1}(1-\tau_k)^{n_k}
\end{equation}
is the probability that the channel is busy in a random slot. $p_b$ can be easily measured by the AP. Through the above two equations, the AP can solve the probabilities $\tau_i$ and $p_i$ numerically. 

In a generic time slot, the probability that node $v$ successfully transmits a packet to the AP is
\begin{align}
\nonumber p_{s,i} =&\; \tau_i(1-\tau_i)^{n_i-1}\prod^{c}_{k=1,k\neq i}(1-\tau_k)^{n_k}\\
            =&\; \frac{\tau_i}{1-\tau_i}(1-p_b).
\end{align}
Therefore, the percentage of resource sharing of node $v$ (which is also the probability that a successful transmission to the AP is from this node) is given by
\begin{align}\label{eq:si}
 s_i =&\; \frac{p_{s,i}}{\sum^{c}_{j=1}n_jp_{s,j}}= \frac{\frac{\tau_i}{1-\tau_i}}{\sum^{c}_{j=1}\frac{n_j\tau_j}{1-\tau_j}}.
\end{align}

The average number of packets (from any of the nodes) received by the AP in one slot is
\begin{align}
\nonumber  \eta =&\; \frac{\textrm{Probability of a successful transmission}}{\textrm{Average length of a slot time}}\\
         =&\; \frac{p_s}{1-p_b+p_sT_s+(p_b-p_s)T_c},
\end{align}
where $p_s=\sum^{c}_{i=1}p_{s,i}$ is the probability of a successful transmission, while $p_b-p_s$ is the probability of a collided transmission. $1-p_b$ is the channel idle probability (i.e., the probability that none of the nodes transmits). $T_s$ and $T_c$ are the numbers of empty slots (i.e., aSlotTime as specified in the standard) of a successful transmission and a collision, respectively. In the case of basic access (without RTS/CTS handshaking), we have \cite{kosek2011simple}
\begin{align}
    T_s =&\; \textrm{AIFSN}_{\min}+L+2SIFS+ACK+2\delta,\\
    T_c =&\; \textrm{AIFSN}_{\min}+L+SIFS+ACK+\delta,
\end{align}
where $L$ is the length of a packet including the MAC and PHY headers\footnote{We assume all the packets are of the same length. Please refer to \cite{bianchi2000performance} for the case with diverse packet lengths.}. $SIFS$ and $ACK$ are durations of a short inter-frame space and an ACK transmission period, respectively. $\delta$ represents the propagation delay. The units of both $T_s$ and $T_c$ are numbers of empty slots. For the cases with RTS/CTS access mechanism, refer to \cite{kosek2011simple} for the derivation of the corresponding $T_s$ and $T_c$. Then, the average number of empty slots between two successive transmissions can be given by
\begin{align}\label{eq:T}
    T = \frac{1}{\eta},
\end{align}
which also describes the frequency of packet arrivals at the AP. Note that, if each misbehaving node is treated as a distinct priority class, the above is able to accommodate both normal and misbehaving nodes.


\section{Hyrid-Share Detector Design}\label{sec:hsdetector}

For a target node belonging to priority class $i$, the hybrid-share detector (called HS detector) is designed as follows. Due to the high nonlinearity of (\ref{eq:taui}) and (\ref{eq:pi}), the numerical solution of $s_i$ may introduce some error, say $\epsilon_i$. Let $\bar{s}_i$ be the numerical solution of (\ref{eq:si}), then
\begin{equation}\label{eq:si}
    s_i=\bar{s}_i+\epsilon_i.
\end{equation}

In the sequel, we shall omit the subscript $i$ since the context is clear. The detector maintains a state $X_k$ with initial state $X_0=0$. Once a packet arrives at the AP, the detector state is updated according to
\begin{equation}\label{eq:hsdetector}
    X_{k+1} = \left[X_k+(I_k-\bar{s})\right]^+,
\end{equation}
where $I_k$ is defined below Eq. (\ref{eq:fsdetector}) and $\mathbb{P}[I_k=1] = s$. Therefore, in normal cases, $X_k$ is expected to remain in $[0,1]$. We introduce a new detection threshold $h$ and make the decision that whether the target node is misbehaving or not by computing
\begin{equation}
    \delta_k = \left\{\begin{array}{ll}
        1, \quad & \textrm{if}~X_k \geq h,\\
        0, \quad & \textrm{otherwise}.
    \end{array}\right.
\end{equation}
Similar as above, $\delta_k=1$ indicates misbehaving.

Note that, in normal cases, $X_k$ may be able to hit $1$ if the AP receives a packet from the target node. Therefore, for the sake of correct detection, $h$ can be set to larger than $1$. For detecting real-time misbehavior of the target node, $X_k$ is reset to $0$ once it hits the threshold $h$. If there is only one access class, $s=\frac1n$, and the HS detector reduces to an FS detector.

We call $X_k$ as the state of the HS detector in step $k$. Note that the step size may vary from time to time because the packet arrivals at the AP are generally random. However, from (\ref{eq:T}) we can obtain the average step size as $T$.

When applying the proposed detector, the AP only needs to compute the MAC model and calculate the percentage of resource sharing once, as long as the network configuration and MAC parameters assigned to each node do not change. As shown in (\ref{eq:hsdetector}), the computation complexity of the proposed detector itself is very low. Therefore, it is worth noting that the proposed detector is light-weight. Moreover, since all received packets are utilized by the detector, misbehavior can be detected in a real-time manner.

\begin{definition}\label{def:falseposrate}
To evaluate the performance of the HS detector, we define the following metrics.
\begin{itemize}
\item The \emph{false positive rate} $p_f$ of the HS detector is the conditional probability that the target node is indicated misbehaving (i.e., $X_k$ is no less than the threshold $h$) when in fact none of the nodes is misbehaving.
\item The \emph{detection rate} $p_d(D)$ of the HS detector is the probability that a misbehaving node will be detected in $D$ time slots (empty slots defined in IEEE 802.11e).
\end{itemize}
\end{definition}

$p_f$ can be viewed as the rate of false alarms, while $p_d(D)$ reflects the effectiveness and real-time performance of the HS detector. Below we analytically analyze the detector performance by modeling $p_f$ and $p_d(D)$.

\subsection{False positive rate}
Without loss of generality, suppose there exists $\sigma>0$ such that both $\frac{\bar{s}}{\sigma}$ and $\frac{1-\bar{s}}{\sigma}$ are integers (say $L_0$ and $L_1$, respectively). For example, we can use the precision of $\bar{s}$ to determine the above two integers. For any step $k$ between two adjacent detector state resettings, suppose there are $k_1$ times that $I_{\kappa}=1$ and $k_0$ times that $I_{\kappa}=0$, where ${\kappa}$ is between the last resetting step and $k$. Thus, based on (\ref{eq:hsdetector}), $X_k\in\{0,X_{k-1}-\bar{s}, X_{k-1}+1-\bar{s}\}$. Furthermore, $X_k \leq k_1 (1-\bar{s})$ which yields that
\begin{align}
    X_k \in \{0,\sigma,2\sigma,\ldots,k_1L_1\sigma\}.
\end{align}
Since $X_k$ is multiples of $\sigma$, its largest possible value is $\bar{m}\sigma$ where $\bar{m}=\lceil\frac{h}{\sigma}\rceil$ (otherwise $X_k$ is reset). Therefore, the support of $X_k$ can be denoted as
\begin{align}
\nonumber  \mathcal{M} =&\; \Bigl\{0,m_1\sigma,m_2\sigma,\ldots,\bar{m}\sigma
                            \Bigl|m_j\in\mathbb{N}^+,m_j< \bar{m}\Bigl\}\\
                    \subseteq&\; \{0,\sigma,2\sigma,\ldots,\bar{m}\sigma\}.
\end{align}
Clearly, $\mathcal{M}$ is a finite set.

According to (\ref{eq:hsdetector}), $X_{k+1}$ depends only on $X_k$ and thus the sequence $\{X_k\}$ forms a homogeneous Markov chain. Since the support of $X_k$ may vary from step to step, to calculate the probabilities of the chain's states at any step $k$, we can consider the bigger set $\{0,\sigma,2\sigma,\ldots,\bar{m}\sigma\}$ without loss of generality. Define
\begin{equation}\label{eq:Xi}
    \boldsymbol{x}_k=\left[\mathbb{P}[X_k=0],\mathbb{P}[X_k=\sigma],\ldots,\mathbb{P}[X_k=\bar{m}\sigma]\right].
\end{equation}
By definition, we have $\boldsymbol{x}_k'*\boldsymbol1=1$, where $\boldsymbol1$ is a vector with all elements equal to 1. Due to the homogeneity of the chain, we can have $\boldsymbol{x}_{k+1}=\boldsymbol{x}_k \mathbf{P}$, where $\mathbf{P}$ is the step-independent probability transition matrix. $\mathbf{P}$ depends only on $s$ and can be also represented as $\mathbf{P}(s)$. Let $P_{i,j}$ be the $(i,j)$th entry of $\mathbf{P}$. To describe $\mathbf{P}$, let us consider the steady-state probabilities of the chain $\{X_k\}$: $\boldsymbol\pi=\lim_{k\to\infty}\boldsymbol{x}_k=[\pi_0,\pi_1,\ldots,\pi_{\bar{m}}]$. Apparently, $\boldsymbol\pi =\boldsymbol\pi \mathbf{P}$. $\pi_m$ can be calculated by considering the following scenarios:
\begin{itemize}
\item If $m=0$, we have $X_k=0$ which happens either if $X_{k-1}\leq \bar{s}$ and $I_{k-1}=0$ (i.e., the received packet is not from the target node) or $X_{k}=\bar{m}\sigma$ and the state is reset subsequently. Therefore,
\begin{align}
 \pi_0 =& \sum^{L_0}_{i=0}\pi_{i}(1-s) + \pi_{\bar{m}},
\end{align}
which indicates that $P_{i,0}=1-s, \forall i\leq L_0$ and $P_{\bar{m},0}=1$.

\item If $0<m\leq L_1$, we have $X_{k-1}=(m+L_0)\sigma$ if $I_k=0$. Hence
\begin{align}
 \pi_m = &\; \pi_{m+L_0}(1-s).
\end{align}
i.e., $P_{m+L_0,m}=1-s$.

\item If $0<m<\bar{m}-L_0$, we have $X_{k-1}=(m+L_0)\sigma$ if $I_k=0$ and $X_{k-1}=(m-L_1)\sigma\}$ otherwise. Hence
\begin{align}
\nonumber \pi_m =&\; \pi_{m+L_0}(1-s) + \pi_{m-L_1}s \\
          =&\; \pi_{m+L_0}P_{m+L_0,m} + \pi_{m-L_1}P_{m-L_1,m}.
\end{align}

\item Otherwise if $\bar{m}-L_0\leq m < \bar{m}$, we do not have the case $I_k=0$. Hence,
\begin{align}
    \pi_m =\pi_{m-L_1}s = \pi_{m-L_1}P_{m-L_1,m}.
\end{align}

\item Finally, when $m = \bar{m}$,
\begin{align}
    \pi_{\bar{m}} = \sum^{L_1}_{i=1} \pi_{\bar{m}-i}s = \sum^{L_1}_{i=1}\pi_{\bar{m}-i}P_{\bar{m}-i,\bar{m}}.
\end{align}
\end{itemize}

Solving these equations, we can get a unique $\boldsymbol\pi$. Based on Definition \ref{def:falseposrate}, the false positive rate is given by
\begin{align}
    p_f = \pi_{\bar{m}}.
\end{align}

\subsection{Average Detection Rate}
Suppose the target node starts to misbehave from step $0$ on and assume that the associated Markov chain $\{X_k\}$ for the normal case before step $0$ has reached its steady state $\boldsymbol\pi$. Note that, with the target node misbehaving (i.e., using different $\textrm{CWmin}$ and/or $\textrm{AIFSN}$), the MAC model changes. Hence, we add superscript $^*$ to the variables defined in previous sections to distinguish the case that the target node is misbehaving from the normal case. Since whether the target node misbehaves or not is not pre-known to the detector, it shall assume that the target node is well-behaving and still use $\bar{s}$ to update its state. Thus, the support of $X_k$ remains the same as above. The only difference lies in the probability of $I_k=1$, which in turn changes the probability transition matrix from $\mathbf{P}(s)$ to $\mathbf{P}^*=\mathbf{P}(s^*)$.


Then, starting at ${{\boldsymbol{x}}}_0={\boldsymbol\pi}$, the Markov chain associated with the $\{X_k\}$ evolves with ${\boldsymbol{x}}_{k+1}={\boldsymbol{x}}_k{\mathbf{P}^*}$. By definition, the average detection rate in time $D$ can be calculated as
\begin{align}
\nonumber  p_d(D) =&\; 1-\prod^{\lfloor\frac{D}{{T}^*}\rfloor}_{k=1}\left(1-\mathbb{P}[X_{k}={\bar{m}}{{\sigma}}]\right)\\
           =&\; 1-\prod^{\lfloor\frac{D}{{T}^*}\rfloor}_{k=1}\left(1-x_{\bar{m},k}\right),
\end{align}
where $\lfloor\frac{D}{{T}^*}\rfloor$ is the average number of steps in time $D$ and $x_{\bar{m},k}$ is the last element of ${\boldsymbol{x}}_k$.

\section{Performance Evaluation}\label{sec:numerical}
Consider a WLAN with one AP and 15 nodes locating close to each other so that they can hear each other's transmissions. The nodes are divided into three priority classes. In class 1, there are $n_1=6$ nodes each of which uses MAC parameters $W_{1}=15$, $\textrm{CWmax}_{1}=1023$ and AIFSN$_1=7$. For the other two classes, we set $n_2=6$, $W_{2}=15$, AIFSN$_2=3$, $n_3=3$, $W_{3}=7$, and AIFSN$_3=2$. $\textrm{CWmax}$ is fixed at 1023 for all the nodes. There is one node (the target node) in class 2 misbehaves.

\begin{figure}[htbp]
  \centering
  \includegraphics[width=2.3in]{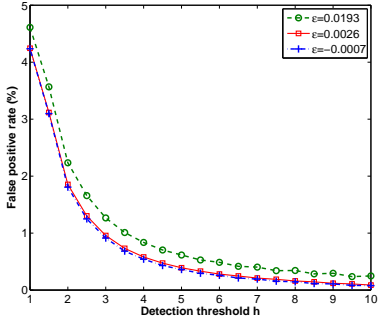}
\vspace{-3mm}
  \caption{False positive rate $p_f$ under various detection thresholds. In the figure, the error $\epsilon=0.0139,0.0026$ and -0.0007 correspond to that $\sigma=\frac{1}{10},\frac{1}{60}$ and $\frac{1}{100}$, respectively.}\label{fig:pf}
\vspace{-4mm}
\end{figure}

To evaluate the false positive rate $p_f$ of the HS detector, we consider the case that the target node well-behaves. As shown in Fig. \ref{fig:pf}, $p_f$ decreases as the detection threshold $h$ increases. This is simply because the higher $h$ is, the less opportunity that the detector state $X_k$ will hit its maximal value (i.e., $\bar{m}\sigma$ as in (\ref{eq:Xi})). The figure also shows that the numerical solution error of the analytical model, as indicated by $\epsilon$ in (\ref{eq:si}), has an impact on the rate $p_f$: a smaller error can result in lower false positive rate. However, since the second and third curves are very close, we can see that a precision of $\frac{1}{60}$ is enough to deliver satisfactory results.

\begin{figure}[htbp]
  \centering
        \subfigure[Under various misbehavior intensities.]{
            \includegraphics*[width=2.3in]{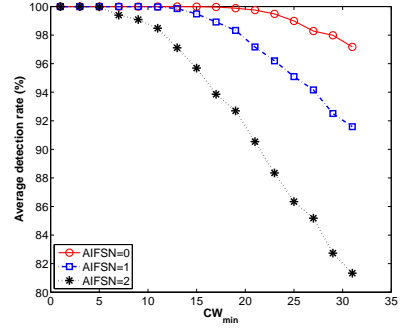}
            \label{fig:detectrateCW}
        \vspace{1mm}
        }
        \subfigure[With different detection window $D$.]{
            \includegraphics*[width=2.3in]{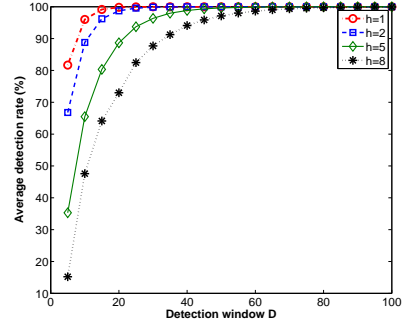}
            \label{fig:detectrateD}
        }
\vspace{-1mm}
\caption{Average detection rate $p_d(D)$.}\label{fig:pd}
\vspace{-4mm}
\end{figure}

We then evaluate the detection rate of the HS detector by considering the misbehaving node with various misbehaving strategies. Fig. \ref{fig:detectrateCW} shows the average detection rate $p_d(D)$ under different misbehaving intensities, where we fix $D=100$ and $h=2.5$. As the misbehavior is intensified (i.e., the target node uses a smaller AIFSN and/or CWmin), more received packets are from the target node. Hence, the detector state increases more frequently and is more likely to hit its maximal value. As a result, the average detection rate increases, which is clearly depicted in this figure.

Fig. \ref{fig:detectrateD} shows the performance of the HS detector associated with the target node under different $D$ and $h$, where the misbehaving strategy is $\textrm{CWmin}=4$ and $\textrm{AIFSN}=0$. We can see that, in all cases, the detector becomes more reliable with a higher detection rate as the detection window gets longer. The misbehavior will be captured almost surely when $D$ is larger than 80. However, a larger $D$ indicates a longer detection delay. In this sense, we should keep $D$ small in order to detect real-time misbehavior. For the similar reason as we discussed above about Fig. \ref{fig:pf}, the higher the detection threshold is, the lower the average detection rate will be achieved. However, since $p_f$ and $p_d(D)$ are two conflict objectives, this figure suggests to carefully choose $h$ and $D$ to balance them.

\section{Conclusion}\label{sec:conclusion}
We have investigated the problem of misbehavior detection in IEEE 802.11e based networks where the nodes are able to choose different priority levels and different MAC parameters. We presented a mathematical model of the percentage of resource sharing of each node, based on which we proposed a hybrid-share detector. Theoretical performance of the detector in terms of false positive rate and average detection rate had been analyzed. Through numerical results, we demonstrated that the false positive rate is sensitive to the detection threshold but tolerable to the error involved in computing the MAC model. The results also indicate that our analysis can help choose proper detection threshold and window to meet real-time requirement while balancing false positive rate and average detection rate.

\section*{Acknowledgement}
This work was supported in part by the NSF under grants CNS-1117687 and CNS-1320736.

\bibliographystyle{IEEEtran}
\bibliography{IEEEabrv,MBD}

\end{document}